\documentclass[11pt]{article}

\usepackage[breaklinks]{hyperref}
\usepackage{graphicx}
\usepackage{tabularx}
\usepackage{amsfonts}
\usepackage{amssymb}
\usepackage{amsmath}
\usepackage{algorithmic}
\usepackage{algorithm}
\usepackage[sort]{natbib}
\bibpunct{(}{)}{;}{a}{,}{,}

\usepackage{color}
\usepackage[normalem]{ulem}

\newcommand{\btheta}{\boldsymbol{\theta}}

\newcommand{\norm}[1]{\left|\left|#1\right|\right|}
\newcommand{\R}{\mathbb{R}}

\title{Squeeze-and-Breathe Evolutionary Monte Carlo Optimisation with Local Search 
  Acceleration and its application to parameter fitting}
\author{Mariano Beguerisse-D\'iaz\,$^{1,3}$\footnote{m.beguerisse-diaz08@imperial.ac.uk}, 
  Baojun Wang\,$^2$, 
  Radhika Desikan\,$^{1}$,\\
  and Mauricio Barahona\,$^{2, 3}$\footnote{m.barahona@imperial.ac.uk}\\
  $^{1}$Department of Life Sciences, $^{2}$Department of Mathematics\\
  Imperial College London. London SW7 2AZ, U.K.
}

\date{\today}

\begin{document}

\maketitle

\begin{abstract}
  \noindent{\bf Motivation:} Estimating parameters from data is a 
  key stage of the modelling process, particularly in biological systems where 
  many parameters need to be estimated from sparse and noisy data sets. 
  Over the years, a variety of heuristics have been proposed to solve this complex 
  optimisation problem, with good results in some cases yet with limitations in the 
  biological setting.\\
  {\bf Results: }In this work, we develop an algorithm for model parameter 
  fitting that combines ideas from evolutionary algorithms, sequential Monte Carlo and direct search optimisation. Our method performs well even when the order of magnitude and/or the range 
  of the parameters is unknown.  The method refines iteratively a sequence of 
  parameter distributions through local optimisation combined with 
  partial resampling from a historical prior defined over the support of all previous 
  iterations. We exemplify our method with biological models using both 
  simulated and real experimental data and estimate the parameters efficiently even in the absence of \textit{a priori} knowledge about the parameters.\\
  {\bf Availability:} Matlab code available from the authors upon request.
\end{abstract}

\section{Introduction}

The increasing drive towards quantitative technologies in Biology has 
brought with it a renewed interest in the modeling of biological systems. 
Models of biological systems and other complex phenomena are generally 
nonlinear with uncertain parameters, many of which are often
unknown and/or unmeasurable \citep{Edelstein-Keshet1988, Alon2007}. 
Crucially,  the values of the parameters dictate 
not only the quantitative but also the qualitative  
behaviour of such models \citep{Strogatz1994, Brown2003}.
A fundamental task in quantitative and systems biology is to use experimental data 
to infer parameter values that minimise the 
discrepancy  between the behaviour of the model and experimental observations. 
The parameters thus obtained can then be cross-validated against unused data 
before employing the fitted model as a predictive tool \citep{Alon2007}.
Ideally, this process could help close the modelling-experiment loop by: 
suggesting specific experimental measurements; identifying 
relevant parameters to be measured; or discriminating between 
alternative models \citep{Gutenkunst2007, Toni2009, Yates2001}.

The problem of parameter estimation and data fitting is classically posed as 
the minimisation of a cost function (i.e., the error) \citep{Gershenfeld1999}. 
In the case of overdetermined linear systems with quadratic error 
functions, this problem leads to least-square solutions, convex optimisations 
that can be solved efficiently and globally based on the 
singular value decomposition of the covariance matrix of the 
data \citep{Lawson1995}. However,  data fitting in nonlinear systems with small 
amounts of data remains difficult, as it usually leads to non-convex optimisations with many local minima \citep{Brewer2008}.  

A classic case in biological modeling is the description of the time evolution 
of a system through ordinary differential equations (ODEs), usually based on mechanistic 
functional forms. Examples include models of biochemical reactions, infectious 
spread and neuronal dynamics \citep{Anderson1992, Edelstein-Keshet1988}. 
Typically, optimal parameters of the nonlinear ODEs must 
be inferred from experimental time courses but the associated optimisation is far from 
straightforward.  Standard optimisation techniques that require an explicit
cost function are unsuitable for this problem due to the difficulty to obtain full analytical 
solutions for nonlinear ODEs \citep{Brown2003, Chen2010, Papachristodoulou2007}.  
Spline-based methods, which approximate the solution though an 
implicit integration of the differential equation \citep{Brewer2008},  
require linearity in the parameters and are therefore not applicable to 
models with nonlinear parameter dependencies, e.g. Michaelis-Menten and Hill 
kinetics.  

Implicit techniques, such as direct search 
methods \citep{Powell1997}, Simulated 
Annealing \citep{Kirkpatrick1983}, Evolutionary 
Algorithms \citep{Mitchell1997,Runarsson2000} 
or Sequential Monte Carlo \citep{Sisson2007}, do not require an explicit cost function. 
However, if as is usually the case, the cost function is a complicated (hyper)surface in parameter space with many local minima, 
gradient and direct search methods tend to get trapped in local minima due to 
their use of local information. 
Although still a local method, 
Simulated Annealing alleviates some of the 
problems related to local minima through the use of stochasticity. However, this 
comes at the cost of high computational overhead and slow convergence and, yet, 
with no guarantee of finding the global minimum.

Instead of an optimisation based on local criteria,
Evolutionary Algorithms (EA) produce an ensemble of possible answers and evolve 
them globally through random mutation and cross-over followed by ranking and 
culling of the worst solutions \citep{Mitchell1997,Runarsson2000, Schwefel1995}. 
This heuristic has been shown to provide an efficient protocol for 
parameter fitting in the life sciences \citep{Moles2003, Zi2006}. However, 
EA methods can be inefficient when the feasible region in 
parameter space is too large, a case typical of models with large uncertainty in 
the parameters.  

Probabilistic methods, such as Sequential Monte-Carlo (SMC) \citep{Sisson2007}, 
propose a different conceptual framework.
Rather than finding a \textit{unique} optimal parameter set, SMC maps 
a prior probability distribution of the parameters 
onto a posterior constructed from samples with low errors until reaching a 
converged posterior.
Recently, SMC has been combined with  Approximate Bayesian Computation 
(ABC) and applied to data fitting and model selection \citep{Toni2009a}. 
However, methods such as ABC-SMC are not only computationally expensive
but also require that the starting prior include the 
\textit{true} value of the parameters. This requirement dents its applicability to 
many biological models, in which not even the order of magnitude
of the parameters is known. In that case, the support of the starting priors must 
be made overly large (leading to extremely slow convergence) in order to avoid the risk 
of excluding the true parameter value from the search space.

In this work, we present an optimisation algorithm for data fitting that takes inspiration from
EA, SMC and direct search optimisation. 
Our method iterates and refines samples from a probability distribution of the 
parameters in a 'squeeze-and-breathe' sequence. At each iteration the probability distribution is 
`squeezed' by the consecutive application of local optimisation
followed by ranking and culling of the local optima.  
The parameter distribution  is then allowed to `breathe' through a random update 
from a historical prior that includes the union of all past supports of the solutions 
(Fig.~\ref{fig:BPM-Panel}). This iteration
proceeds until convergence of the distribution of solutions and their average error. 
A key feature of the algorithm is the accelerated 
step-to-step convergence through a combination of local optimisation and of 
culling of local solutions. Importantly, the method can also find parameters that 
lie outside of the range of the initial prior, and can deal with parameter values 
that extend across several orders of magnitude. We now provide definitions 
and a full description of our algorithm and showcase its applicability to different 
biological models of interest.

\section{Algorithm}
\subsection{Formulation of the problem}
\label{sec:formulation}

Let $\mathbf{X}(t) = [x_1(t), \dots, x_d(t)]$ denote the state of a 
system with $d$ variables at time $t$.
The time evolution of the state is described by a system of (possibly nonlinear) ODEs:
\begin{equation}
  \dot{\mathbf{X}} = f(\mathbf{X}, t; \, \btheta).
  \label{eq:syst}
\end{equation}
Here \mbox{$\btheta = [\theta_1, \dots, \theta_N]$} is the vector of $N$ 
parameters of our model. 

The experimental data set is formed by $M$ observations of some of the variables of 
the system:
\begin{equation}
  \mathcal{D} = \left\{\widetilde{\mathbf{X}}(t_i)\,|\, 
  i = 1, \dots, M \right \}.
  \label{eq:data}
\end{equation}
Ideally, $M > 2N+1$ since $2N+1$ experiments are enough for unequivocal 
identification of an ODE model with $N$ parameters when no measurement error is 
present \citep{Sontag2002}.

The {\it cost function} (i.e., the error) 
to be minimised is:
\begin{equation}
  E_{\mathcal{D}}(\btheta) = \sum_{i=1}^{M}\norm{\mathbf{X}(t_i; \btheta)
 - \widetilde{\mathbf{X}}(t_i)},
  \label{eq:obj-fun}
\end{equation}
where $\norm{\cdot}$ is a relevant vector norm. 
A standard  choice is the 
Euclidean norm (or 2-norm) which corresponds to the sum of squared errors:
 \begin{equation}
  E^{(2)}_{\mathcal{D}}(\btheta) = \sum_{i=1}^{M} \sum_{j=1}^{d'} \left({X}_j(t_i;
 \btheta) -\widetilde{X}_j(t_i) \right)^2,
  \label{eq:obj-fun-euclidean}
\end{equation}
where we assume that $d'$ variables are observed.
The cost function \mbox{$E_{\mathcal{D}}:\R^N \to \R_{+}$} 
maps a $N$-dimensional parameter vector onto its corresponding error, 
thus quantifying how far the data and the model predictions are for that 
particular parameter set. 

The aim of the data fitting procedure is to find the parameter vector 
$\btheta^{**}$ that minimises the error globally subject to restrictions dictated 
by the problem of interest:
\begin{equation}
  \btheta^{**} = \underset{\btheta}{\min} \,  E_{\mathcal{D}}(\btheta), \quad
   \text{subject to constraints on $\btheta$.}
 \label{eq:min-prob}
\end{equation}

\subsection{Definitions}
\label{sec:definitions}

\begin{itemize}

\item \textit{Data set:} $\mathcal{D}$, a set of $M$ observations, as defined 
  in Eq.~(\ref{eq:data}). 

\item \textit{Parameter set:} $\btheta = [\theta_1,\dots , \theta_N] \in \R^N_{+}$.  
  Due to the nature of the models considered, $\theta_i \geq 0, \,\, \forall i$. 
  
\item \textit{Objective function:} $E_{\mathcal{D}}(\btheta)$,  the error function 
to be minimised, as defined in Eq.~(\ref{eq:obj-fun-euclidean}).  

\item \textit{ Set of local minima of \mbox{$E_{\mathcal{D}}(\btheta)$}: } 
$ \mathbb{M} = \{\btheta^*~|~E_{\mathcal{D}}(\btheta^*)~\leq~ E_{\mathcal{D}}(\btheta), \break \forall\btheta \in \mathcal{N}(\btheta^*)\}$ 
where $\mathcal{N}(\btheta^*)$ is a neighbourhood of $\btheta^*$. 

\item \textit{Global minimum of \mbox{$E_{\mathcal{D}}(\btheta)$}:} 
 $\btheta^{**}$, a parameter set such that $E_{\mathcal{D}}(\btheta^{**}) \leq   E_{\mathcal{D}}(\btheta)$, $\forall \btheta$. Clearly, $\btheta^{**} \in \mathbb{M}$.

\item \textit{Local minimisation mapping:} \mbox{$L:\R^N_{+} \to \mathbb{M}$}. 
  Local minimisation maps  $\btheta$  onto a local minimum:
  \mbox{$L(\btheta)=\btheta^* \in \mathbb{M}$}. 
  
\item \textit{Ranking and culling of local minima}: 
  $\{ \btheta^\dagger \} _{1}^{B} = \mathcal{RC}_ B \left (\{ \btheta \} _{1}^{J} \right )$.
  This operation ranks $J$ parameter sets and selects the $B$ parameter sets with the 
  lowest $E_{\mathcal{D}}$. 

\item \textit{Joint probability distributions of the parameters at iteration $k$:}  $\pi_k(\btheta)$ (prior) and  $\varpi_k(\btheta)$ (posterior).

\item \textit{Marginal probability distribution of the $i^{th}$ component 
  of $\btheta$}:  For instance, $\pi(\theta_i) = \int \pi(\btheta) \, \prod_{r \neq i} d\theta_r.$

\item \textit{Historical prior at iteration $k$:}  
$\zeta_k(\btheta)=\prod_{i=1}^N\zeta_k(\theta_i)$ where
\begin{equation}
  \zeta_k(\theta_i) \sim U \left( \min \left(  \mathfrak{Z}_k(\theta_i) \right ),
  \max \left( \mathfrak{Z}_k(\theta_i) \right) \right).  
  \label{eq:hist-prior-unif}
\end{equation}
Here $U(a,b)$ is a uniform distribution with support in $[a,b]$ and 
$\mathfrak{Z}_k(\theta_i) = \zeta_{k-1}^{-1} \cup \varpi^{-1}_k$ is the 
union of the supports of $\varpi_k(\theta_i)$ and $\zeta_{k-1}(\theta_i)$.
  
\item \textit{Update of the prior at iteration $k$:} $\pi_k(\btheta)=\prod_{i=1}^N\pi_k(\theta_i)$ with
\begin{equation}
  \pi_k(\theta_i) \sim p_m  \varpi_k(\theta_i) + (1-p_m) \zeta_k(\theta_i),
  \label{eq:update-prior}
\end{equation}
that is,  a convex mixture of the posterior and the historical prior with weight $p_m$.

\item {\it Re-population:} Obtain population of $J$ random points simulated 
from the prior $\pi_{k-1}(\btheta)$. 

\item \textit{Convergence criterion for the error:} The difference between the means of the errors of the posteriors in consecutive iterations is smaller than the pre-determined tolerance:
\begin{equation}  
\phi_k = \overline{E_{\mathcal{D}}(\varpi_{k-1}(\btheta))}-\overline{E_{\mathcal{D}}(\varpi_k(\btheta))} < Tol.
\end{equation}
  
\item \textit{Convergence criterion for the empirical distributions:} 
  The samples of the posteriors in consecutive iterations
are indistinguishable at the 5\% significance level according to the nonparametric  
Mann-Whitney rank sum test:  
\begin{equation}
\mathcal{MW}\left(\varpi_k(\btheta),\varpi_{k-1}(\btheta) \right) = 0.
\end{equation}   
\end{itemize}

\subsection{Description of the algorithm}
\label{sec:algorithm}

\begin{figure*}[!tpb]
  \centerline{\includegraphics[width=\textwidth]{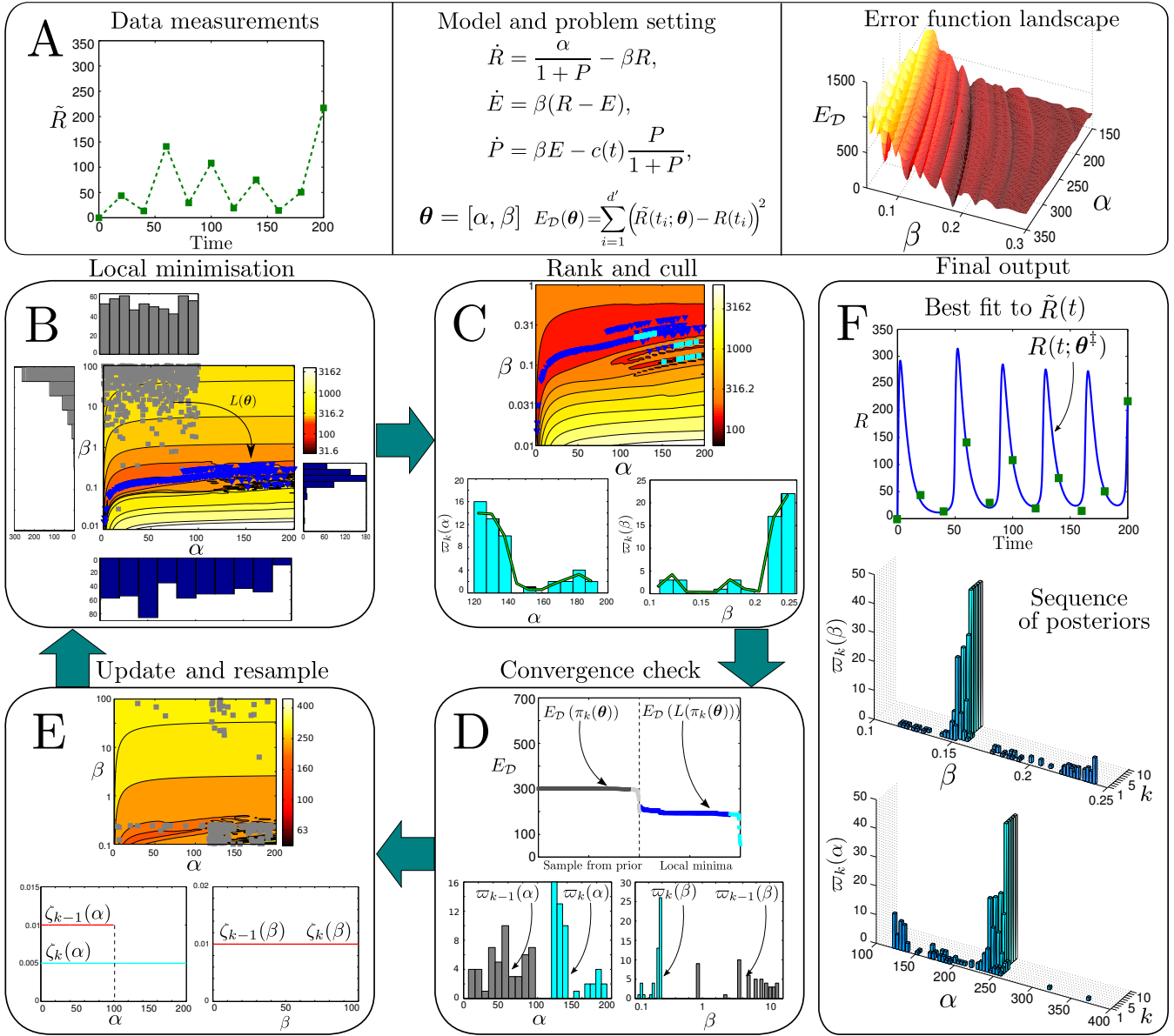}}
  \caption{{\small (Colour online) Steps of Algorithm~\ref{alg:method} exemplified through
    the BPM model~(\ref{eq:BPM-syst}). 
{\bf A}: The problem is defined 
by the data set, the model and the error function to be minimised. Note the rugged landscape of the error function in the parameter plane $(\alpha, \beta)$, with many local minima.
    {\bf B}: In the first iteration, we simulate $J$ points in parameter space from the uniform initial prior $\pi_0(\btheta)$ (grey squares, grey histograms) which are then minimised locally with a Nelder-Mead algorithm $L(\btheta)$ (blue triangles, blue histograms). The local optimisation aligns the parameter sets onto the level curves of $E_{\mathcal{D}}$.
    {\bf C}: The $B$ best local minima (top, light blue squares)
    are selected and considered to be samples from the posterior distribution (bottom, light blue histograms).
    {\bf D}: Convergence of the error of the samples (top, local minima in blue, $B$ lowest in 
    light blue) and of the posterior distributions (bottom, light blue) are 
    checked against the errors of the sample (top, in grey)  and the priors 
    (bottom, in grey).
    {\bf E}: If convergence is not achieved, the historical prior is
    updated (previous historical prior in red updated to light blue) and
    a new set of $J$ points are simulated from the posterior with probability $p_m$ and 
    from the historical prior with probability $1-p_m$ (grey squares).
    This new sample is fed back to the local minimisation step {\bf B} .
    {\bf F}: The algorithm stops when convergence is reached (after nine iterations, in this case) 
providing an optimal parameter set
$\btheta^\ddagger$ (top, time course of optimal model in blue)  and the sequence of 
optimised posteriors at each iteration (bottom).}}
  \label{fig:BPM-Panel}
\end{figure*}

\begin{algorithm}[!tpb]
  \caption{Squeeze-and-Breathe optimisation.}
  \label{alg:method}
  \begin{algorithmic}
  \STATE Set running parameters of algorithm: 
    $B, J\in \mathbb{N}$, $p_m\in [0,1]$, \textit{Tol}
    \STATE Choose initial priors $\pi_0(\btheta)$ and $\zeta_0(\btheta)$.
    \STATE Set $\mathcal{H}_0 = \emptyset $ and $k \gets 1$.
    \REPEAT
      \STATE  Let $\mathcal{H}_k = \mathcal{H}_{k-1}$.
      \STATE Simulate $J$ points from $\pi_{k-1}(\btheta)$ through re-population.
      \FOR{$\ell=1 \to J$} 
         \STATE Obtain local minimum $\btheta_\ell^* = L(\btheta_\ell)$.
         \STATE Store the pair 
         $[\btheta_\ell^*, E_{\mathcal{D}}(\btheta_\ell^*)]$ in $\mathcal{H}_k$.
      \ENDFOR
      \STATE  Rank and cull the set of local minima: 
      $\mathcal{H}_k = \mathcal{RC}_B \left(\mathcal{H}_k \right)$ 
      \STATE Define the posterior  $\varpi_k(\btheta)$ from the sample 
      $\mathcal{H}_k$.  
      \STATE Update $\zeta_k(\btheta)$ from 
      $\zeta_{k-1}(\btheta)$ and $\varpi_k(\btheta)$.
      \STATE Update the prior 
      $\pi_k(\btheta) \sim p_m \varpi_k(\btheta) + (1-p_m)\zeta_k(\btheta)$.
      \STATE $k \gets k+1$.
    \UNTIL{$\phi_k < Tol$ and 
      $\mathcal{MW}\left(\varpi_k(\btheta),\varpi_{k-1}(\btheta) \right) = 0$}
  \end{algorithmic}
\end{algorithm}

Algorithm~\ref{alg:method} presents the pseudo-code for our method using 
the definitions above. The iterations produce progressively more refined 
distributions of the parameter vector. At each iteration~$k$, 
a population simulated from the prior distribution $\pi_{k-1}(\btheta)$ is locally 
minimised followed by ranking 
and culling of the local minima to create a posterior distribution 
$\varpi_{k}(\btheta)$ (squeeze step). This 
distribution is then combined with an encompassing historical prior to generate 
the updated prior $\pi_{k}(\btheta)$ (breathe step). The iteration loop terminates 
when the difference between the mean errors of 
consecutive posteriors is smaller than the tolerance
and the samples of the posteriors are indistinguishable. 
We now explain these steps in detail (Fig.~\ref{fig:BPM-Panel}) through the BPM model 
(see Sec.~\ref{sec:BPM}). 

\begin{enumerate}
\item  \textit{Formulation of the optimisation:}
  The data set $\mathcal{D}$ and the model equations parameterised by $\btheta$ 
  allow us to define an error function $E_{\mathcal{D}}(\btheta)$ whose global minimum corresponds 
  to the best model. 

In our illustrative example, the BPM model~(\ref{eq:BPM-syst}) has the 
parameter vector $\btheta = [\alpha, \beta]$ and the error function is depicted in 
Fig.~\ref{fig:BPM-Panel}A.  The global optimisation on the rugged landscape of this function is computationally  hard.  
\item \textit{Initialisation:}
  \begin{itemize}
  \item Set the running parameters of the algorithm: the size of the simulated  
    population, $J$; the size of the surviving population after 
    culling, $B$; the update probability, $p_m$; and the tolerance,
    $Tol$.

    In this example, $J=500$, $B=50$, $p_m = 0.95$ and $Tol=10^{-5}$.      
    
  \item Choose $\pi_0(\btheta)$, the initial prior distribution of the parameter 
    vector.  

    In this case, we take 
    $\alpha$ and $\beta$ to be independent and uniformly distributed: 
      $\pi_0(\btheta) \sim U(0,100) \times U(0,100)$. 

  \item Initialise $\zeta_0(\btheta) = \pi_0(\btheta)$, the historical prior of 
    the parameters. 
    
  \item Simulate $J$ points from $\pi_{0}(\btheta)$ to generate the 
    initial sample $\{\widehat\btheta_{0}\}_{1}^{J}$. 
  \end{itemize}

\item \textit{Iteration (step $k$):}  Repeated until termination criterion is
  satisfied.  
  Figure~\ref{fig:BPM-Panel} shows the first iteration of our method applied to 
  the BPM example.
  \begin{enumerate}
  \item \textit{Local minimisation:}    
    Apply local minimisation to the simulated parameters from the 'prior'
    $\{\widehat\btheta_{k-1}\}_{1}^{J}$ and map them onto local minima of 
    $E_{\mathcal{D}}(\btheta)$ to generate  
    $\{ L(\widehat \btheta_{k-1}) \}_{1}^J  \in \mathbb{M}$.

    Here we use the Nelder-Mead simplex method \citep{Nelder1965},
    though others can be used.     
    Figure~\ref{fig:BPM-Panel}B shows the simulated points from 
    $\pi_{0}(\btheta)$ (grey squares) and its corresponding histograms (in grey).
    After local minimisation, this sample is mapped onto the dark blue triangles in 
    Fig.~\ref{fig:BPM-Panel}B (histograms in dark blue).
    Note how the local minima align with  the level curves of $E_{\mathcal{D}}$ with 
    a markedly different distribution to the uniform prior. 
    Note also that many of the optimised values of  
    $\alpha$ lie outside the range of the prior $(0,100)$ and are now 
    distributed  over the interval $(0,200)$. On the other hand, the values of $\beta$  
    have collapsed inside $(0,1)$. 

  \item \textit{Ranking and culling:} 
    Rank the $J+B$ local minima from the $k-1$ and $k$ iterations, select the 
    $B$ points with the lowest $E_{\mathcal{D}}$ and cull (discard) the rest: $$\mathcal{RC}_B \left(  \{  L(\widehat \btheta_{k-1}) \} _{1}^{J} \cup \{\widehat\btheta_{k-1}^\dagger\}_{1}^{B}  \right ) = \{\widehat\btheta_k^\dagger\}_{1}^{B}.$$ 
Denote the best parameter vector of this set as $\btheta_k^\ddagger = \underset{E_{\mathcal{D}}}{\min} \left (\{ \widehat\btheta_k^\dagger  \}_{1}^B  \right )$. We consider $\{\widehat\btheta_k^\dagger\}_{1}^{B}$ to be a sample from  the optimised (`posterior') distribution, $\varpi_k(\btheta)$. 

 The $B=50$ best parameter sets are shown (light blue squares) in 
 Fig.~\ref{fig:BPM-Panel}C (light blue histograms).

\item \textit{Termination criterion:} Check that the 
  difference between the mean errors of the consecutive optimised samples 
  is smaller than the tolerance: $\phi_k \leq Tol$. 
  We also gauge the `convergence' of the posteriors through 
   the Mann-Whitney (MW) test to determine if the samples from consecutive 
  posteriors are distinguishable: 
  $$\mathcal{MW} (\varpi_{k-1}(\btheta), \varpi_k(\btheta)) \equiv  \mathcal{MW} \left( \{\widehat\btheta_{k-1}^\dagger\} _{1}^{B}, \{\widehat\btheta_k^\dagger\}_{1}^{B} \right ), $$
  where $\mathcal{MW}$ is a $0$-$1$ flag. The MW test gives additional 
information about the change of the optimised posteriors from one iteration to the next.

Figure~\ref{fig:BPM-Panel}D shows the convergence check for the first 
iteration of the BPM model: (i)~top, errors of the sampled prior (grey, left)
 with errors of the local minima (dark blue, right) and the $B$ surviving points 
 (light blue); (ii)~bottom, histograms of the prior (grey) and the 
posterior (light blue). Clearly, in this iteration neither the error nor the 
distributions have converged so the algorithm does not stop.

\item \textit{Update of historical prior and generation of new sample:}
If convergence is not achieved, update the historical prior $\zeta_k(\btheta)$ 
as a uniform distribution over the union of the supports of the existing historical 
prior and the calculated posterior~(\ref{eq:hist-prior-unif}).
Equivalently, the support of the historical prior extends over 
the union of the sequence of all historical priors
$\{ \zeta_0(\btheta), \ldots, \zeta_{k-1}(\btheta) \}$ and of all posteriors 
$\{ \varpi_1(\btheta), \ldots, \varpi_{k}(\btheta) \}$.

As shown in Fig.~\ref{fig:BPM-Panel}E for the BPM example,  the marginal of the 
historical prior for $\alpha$ is expanded to $U(0,200)$, since the optimised 
parameter sets have reached values as high as 200.  Meanwhile,  the $\beta$
marginal of the historical prior remains unchanged as $U(0,100)$ because there 
has been no expansion of the support.  

The historical prior is used to mutate the updated prior before 
the next iteration by constructing a weighted mixture of the 
posterior and the historical prior with weight $p_m$, as shown in~(\ref{eq:update-prior}).
We re-populate from this updated prior by simulating from the posterior with probability 
$p_m=0.95$ and from the historical prior with probability $(1-p_m)$ to generate 
the new sample $\{\widehat\btheta_{k}\}_{1}^{J}$ and iterate back. 

Figure~\ref{fig:BPM-Panel}E shows the sample of $J$ points simulated from the 
new prior. The $\alpha$-components of most points are between 100 and 200 
and the $\beta$-components are between 0.1 and 1.0, but there are a few
that lie outside the support of the posterior. The process in panels B.~C,~D, 
and~E of Fig.~\ref{fig:BPM-Panel} is iterated for this new set of points. 
\end{enumerate}
\item \textit{Output of the algorithm:} 
  When the convergence criteria have been met, the iteration stops at iteration 
  $k^*$ and the last $\btheta_{k^*}^\ddagger$  is presented as the optimal parameter 
  set for the model. We can also examine the sequence of optimised parameter 
  distributions $\{ \varpi_1(\btheta), \ldots, \varpi_{k^*}(\btheta) \}$ obtained for 
  all iterations (Fig.~\ref{fig:BPM-Panel}F).
\end{enumerate}

\section{Application to biological examples}
\label{sec:examples}

We apply our algorithm to three biological examples of 
interest. The first two correspond to simulated data from models in the literature, 
while in the third  example we apply our algorithm to unpublished experimental  
data of the dynamical response 
of an inducible genetic promoter constructed for an application in Synthetic Biology.

\subsection{BPM model of gene-product regulation}
\label{sec:BPM}

\begin{table}[!t]
     {\begin{tabularx}{170mm}{X|XXXXXX}\hline
         &  Min. &  &  & Conv. & Conv. &   \\ 
         $k$  & Error & $\alpha_k^\ddagger$ & $\beta_k^\ddagger$ & $\varpi_k(\alpha)$
         & $\varpi_k(\beta)$ &  $\phi_k$ \\ \hline
         1  & 56.0941 & 193.7447 & 0.1304 & - & - & - \\
         2  & 28.2735 & 246.7510 & 0.1528 & No & No & 133.9020 \\
         3  & 27.2083 & 248.7557 & 0.1532 & No & No & 6.8542 \\
         4  & 26.9838 & 250.3593 & 0.1536 & No & No & 0.6532 \\
         5  & 26.6504 & 251.7189 & 0.1538 & No & No & 0.3281 \\
         6  & 26.6504 & 251.7189 & 0.1538 & No & No & 0.1963 \\
         7  & 26.6504 & 251.7189 & 0.1538 & Yes & Yes & 0.0118 \\
         8  & 26.6504 & 251.7189 & 0.1538 & No & No & 0.0131 \\
         9  & 26.6504 & 251.7189 & 0.1538 & Yes & Yes & $1.414\times 10^{-6}$ \\
         \hline
     \end{tabularx}}
     \caption{{\small Results of the fitting of the BPM model 
       with Algorithm~\ref{alg:method}: smallest error of iteration $k$; the best values $\alpha_k^\ddagger$ and $\beta_k^\ddagger$; 
       whether the distributions have converged; and the
       difference of the mean errors of the optimised population.}}
     \label{tab:BPM-results}
\end{table}

The Bliss-Painter-Marr (BPM) model \citep{Bliss1982} describes the behaviour of a 
gene-enzyme-product control unit with a negative feedback loop:
\begin{align}
  \dot{R} & = \frac{\alpha}{1 + P} - \beta R, \nonumber \\
  \dot{E} & = \beta(R-E), \label{eq:BPM-syst} \\
  \dot{P} & = \beta E - c(t)\frac{P}{1+P}. \nonumber 
\end{align}
Here, $R, E$ and $P$ are the concentrations (in arbitrary units) of mRNA, enzyme 
and product, respectively.  The degradation rate of the product has an explicit 
time dependence, which in this case has the form of a ramp saturation:
\begin{equation*}
  c(t) = \left\{
  \begin{array}{cl}
    5 + 0.2t & 0 \leq t < 50, \\
    15 & t \geq 50.
  \end{array}
  \right.
\end{equation*}
The model represents a gene that codes for an enzyme which in turn 
catalyses a product that inhibits the transcription of the gene. 
This self-inhibition can lead to oscillations, 
which have been shown to occur in the tryptophan operon in 
{\it E.~coli} \citep{Bliss1982}. 

We construct a data set from simulations of this model with 
$\btheta_{\mathrm{real}} =[\alpha, \beta]= [240, 0.15]$ and initial conditions 
$R(0)=E(0)=P(0)=0$. 
The data set $\mathcal{D}$ consists of 10 measurements of $R(t)$ at 
particular times with added gaussian noise drawn from $\mathcal{N}(0, 15^2)$ 
(Table~\ref{tab:BPM-data}).  The error function 
$E_{\mathcal{D}}(\btheta)$~(\ref{eq:obj-fun-euclidean}) corresponds to a non-convex optimisation 
landscape\footnote{We thank Markus Owen of the University of Nottingham 
for suggesting this example.}: a complex rugged surface with many local minima making global 
optimisation hard (Fig.~\ref{fig:BPM-Panel}A).

We use Algorithm~\ref{alg:method}  to estimate the `unknown' parameter values 
from the `measurements' of $R$, as illustrated in 
Sec.~\ref{sec:algorithm} and Fig.~\ref{fig:BPM-Panel}.
Feigning ignorance of the true values, we choose a uniform prior 
distribution with range $[0,100]$ for both parameters: 
\mbox{$\pi_0 (\btheta) \sim [U(0,100), U(0,100)]$.}  
The rest of the paramters are set to: 
$J=500$, 
$B=50$,  
$p_m = 0.95$  and 
$Tol=10^{-5}$. 
Note that the \textit{true} value of $\alpha$ falls outside of the assumed range 
of our initial prior, while the range of $\beta$  in our initial prior is two orders 
of magnitude larger than its true value.  This level of uncertainty about parameter values is typical 
in data fitting for biological models.

Figure~\ref{fig:BPM-Panel} highlights a key aspect of our algorithm: the local 
minimisation can lead to local minima outside of the range of the initial prior. 
Furthermore, our definition of the historical prior ensures that successive iterations 
can find solutions within the largest hypercube of optimised solutions in parameter 
space.  In this example, the algorithm moves away from the $U(0,100)$ prior 
for $\alpha$ and finds a distribution around 240 (the true value) after three 
iterations, while in the case of $\beta$, the distribution collapses to values 
around 0.15 after one iteration. Although the algorithm finds the minimum 
$\btheta^\ddagger$ after 5 iterations, the algorithm is terminated after 9 iterations, 
when the posterior distributions are similar (according to the MW test) 
and the mean errors have also converged (Table~\ref{tab:BPM-results}).
The estimated parameters for this noisy data set
are $\btheta_{k^*}^\ddagger= [251.7189, 0.1530]$. In fact, the error of
the estimated parameter set is lower than that of the real parameters:
$E_{\mathcal{D}}(\btheta^\ddagger) =26.65 <  E_{\mathcal{D}}(\btheta_{\mathrm{real}}) = 28.26$, 
due to the noise introduced in the data. When a data set without noise
is used, the algorithm finds the true value of the parameters to
9 significant digits (not shown).

\subsection{SIR epidemics model}

\begin{figure*}[!t]
  \centerline{\includegraphics[width=\textwidth]{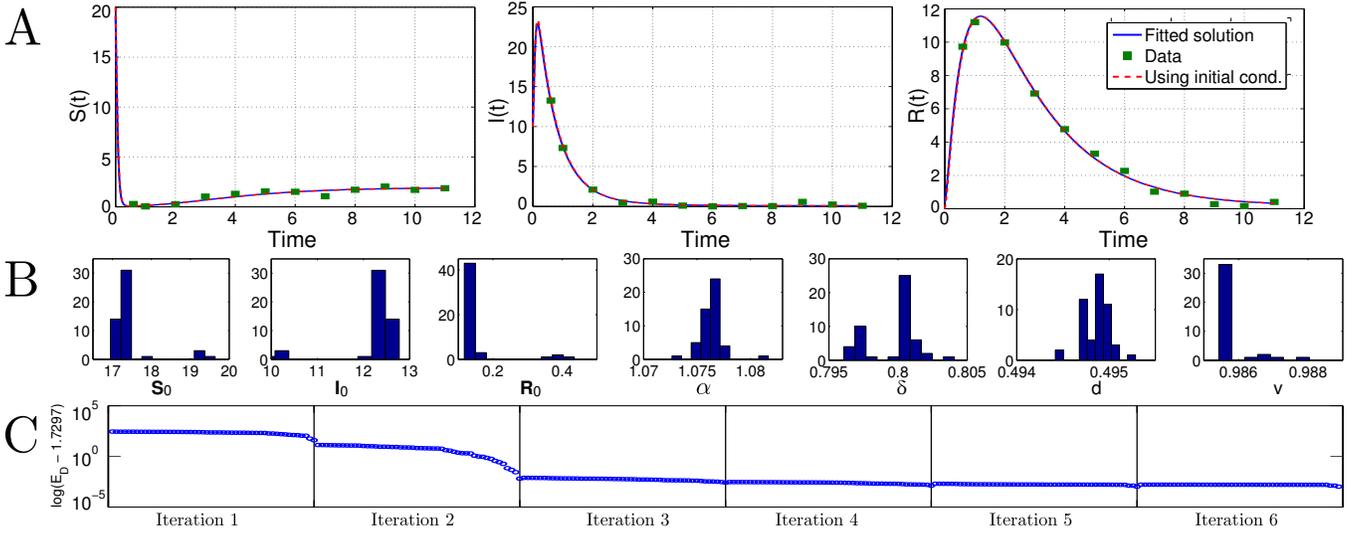}}
  \caption{{\small (Colour online) 
    {\bf A}: Time courses of the SIR model~(\ref{eq:SIR-model}). 
    Green squares 
    are simulated `data' points  (Table~\ref{tab:SIR-data}) and
    bold blue lines are the model fit with the best parameters 
    $\alpha^\ddagger=1.0726$, $\gamma^\ddagger=0.7964$, $d^\ddagger=0.4945$, and 
    $v^\ddagger=0.9863$ and the best fit initial conditions $S_0^\ddagger=19.1591$, 
    $I_0^\ddagger=10.3016$, and $R_0^\ddagger=0.3861$. Red dashed
    lines use the best fit parameters and the real initial conditions. 
    The minimum error is $E_{\mathcal{D}}(\btheta^\ddagger)= 1.7297$.
    {\bf B}: Histogram of the values of the 50 best 
    parameters and initial conditions of the model obtained after convergence at 
    six iterations.
    {\bf C}: Convergence of the error of the optimised samples at every 
    iteration relative to the final error.}}
  \label{fig:SIR-Panel}
\end{figure*}

Susceptible-Infected-Recovered (SIR) models are widely used in epidemiology to 
describe the evolution of an infection in a population \citep{Anderson1992}.  
In its simplest form, the SIR model has three variables: the susceptible population
$S$, the infected population $I$ and the recovered population $R$: 
\begin{align}
  \dot{S} & =  \alpha - (\gamma I + d)S, \nonumber \\
  \dot{I} & =  (\gamma S - v - d)I, \label{eq:SIR-model} \\
  \dot{R} & =  vI - dR. \nonumber
\end{align}
The first equation describes the change in the susceptible population, growing 
with birth rate $\alpha$ and decreasing by the rate of infection $\gamma IS$ and 
the rate of death $dS$. The infected population grows by the rate of 
infection $\gamma IS$ and decreases by the rate of recovery $vI$ and the rate of 
death $dI$. The recovered population grows by the rate of recovery $vI$ 
and decreases by the death rate $dR$. Here we use the same form of the equations 
as \cite{Toni2009a}. 

The data generated from the model~(\ref{eq:SIR-model}) (see Table~\ref{tab:SIR-data}) was obtained directly from \cite{Toni2009a}. Hence the 
original parameter values were not known to us and further we assumed the initial
conditions also to be unknown and fitted them as parameters. 
We used Algorithm~\ref{alg:method} to estimate $\alpha$, $\gamma$, $v$, and $d$ and 
initial conditions $S_0$, $I_0$, and $R_0$. The prior marginal distributions for all 
parameters were set as  $U(0,100)$.
The other parameters were set to: $J=1000$, $B=50$, $p_m=0.95$ and $Tol=10^{-5}$. 
The algorithm converged after six iterations. Figure~\ref{fig:SIR-Panel}A shows 
the prediction of the model~(\ref{eq:SIR-model}) with the best parameters estimated by our algorithm. 
The fit is good with little difference between the curves obtained using the real initial 
conditions and the ones estimated by our method. 

The posterior distributions after six iterations of the algorithm are shown on 
Fig.~\ref{fig:SIR-Panel}B. The errors obtained after each local minimisation in 
a decreasing order on each iteration are shown on a semilogarithmic scale in 
Fig.~\ref{fig:SIR-Panel}C. We can observe how the errors decrease several 
orders of magnitude over the first three iterations and converge steadily during 
the last three iterations until $\phi_k\leq Tol$.

\subsection{An inducible genetic switch from Synthetic Biology}

\begin{figure}[!tbp]
  \centerline{\includegraphics[width=\textwidth]{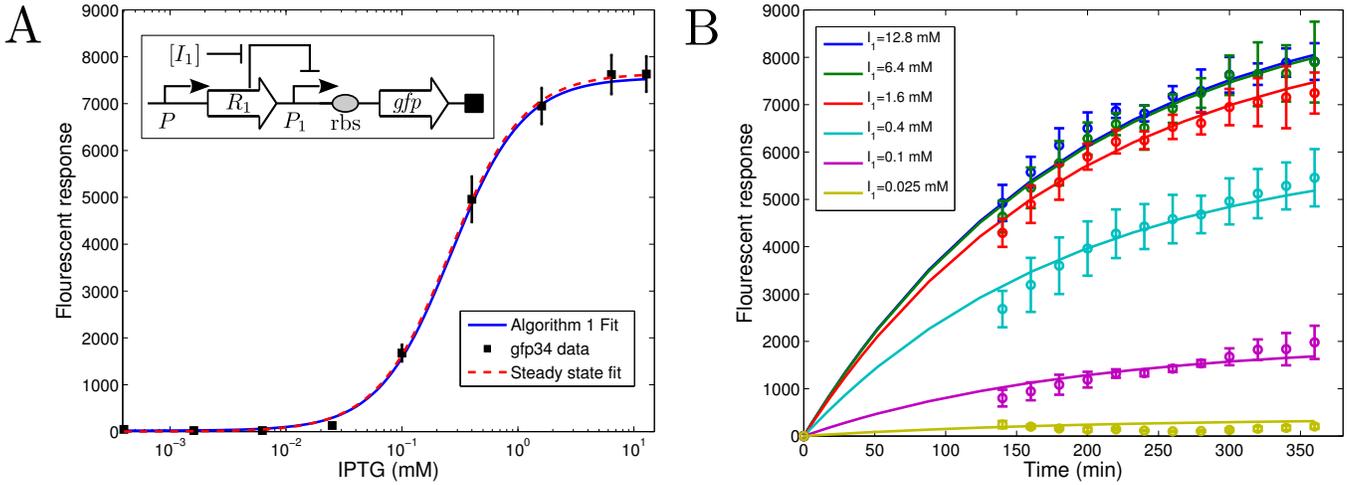}}
  \caption{{\small{\bf A}: (Colour online) 
    {\it Inset}: An inducible genetic switch consisting 
    of $P_1$, a negatively regulated environment-responsive promoter. 
    The repressor R$_1$ promoted by $P$ regulates $P_1$. The switch is responsive to 
    an exogenous inducer $I_1$, which binds to R$_1$ to relieve its repression on 
    $P_1$ and to turn on the transcription of the downstream target gene, such as 
    a {\it gfp}. The ribosome binding site (rbs) is 
    used to tune the translation efficiency of the downstream gene.
    {\it Plot}: Fluorescent response of the switch 
    with {\it gfp}-34 to different doses of IPTG (circles).  Stationary solutions of 
    Eq.~(\ref{eq:GFP}) using the parameters obtained with 
    Algorithm~\ref{alg:method} (solid lines). 
    {\bf B}: Time course of the fluorescent response of the switch with
    {\it gfp}-34 to several doses of IPTG (circles) and time-dependent solutions 
    of Eq.~(\ref{eq:GFP}) using the parameters obtained with Algorithm~\ref{alg:method} (solid lines). 
    Similarly good fits were obtained for responses to \mbox{$I_1$ = 0.0063, 0.0016, 0.0004, and 0.0 mM} (not shown).}}
  \label{fig:GFP-panel}
\end{figure}

The use of inducible genetic switches is widespread in 
synthetic biology and bioengineering as building blocks for more 
complicated gene circuit architectures. An example is shown 
schematically in the inset of Fig.~\ref{fig:GFP-panel}A. 
This environment-responsive switch 
is used to control the expression of a target gene $G$ (usually tagged with green 
fluorescent protein or {\it gfp}) through the addition of an exogenous small 
molecule $I_1$ (e.g., isopropyl thiogalactopyranoside  or IPTG). The input-output 
behaviour of this system can be described by the following ordinary differential 
equation \citep{Szallasi2006, Alon2007}:
\begin{equation}
  \dot{G} = \alpha k_1 + \frac{k_1 I_1^{n_1}}{K_1^{n_1} + I_1^{n_1}} - dG.
  \label{eq:GFP}
\end{equation}
Here, $\alpha k_1$ is the basal activity of the promoter $P_1$ 
and $dG$ is the linear degradation term. 
The second term is a Hill function that models the 
cooperative transcription activation in response to the inducer $I_1$   
with maximum expression rate $k_1$, constant~$K_1$ and Hill coefficient~$n_1$.

The $lacI$--$P_{lac}$ switch has been characterised 
experimentally in response to different doses of 
IPTG in \cite{Wang2010,Wang2011}. Equation~(\ref{eq:GFP}) 
can be solved explicitly and one can use nonlinear least squares and
the analytical solution to fit data at stationarity (i.e., at long times) 
and estimate $\alpha$, $n_1$, $K_1$, and the ratio $k_1/d$. 
These estimates have been obtained assuming equilibrium (\mbox{$\dot{G}=0$}) and 
initial condition $G(0)=0$ by \cite{Wang2011} (Table~\ref{tab:GFP-data}). 

In fact, the experiments measured time series of the expression of $G$ 
every 20 minutes from \mbox{$t=140$ to $360$ min.} for different doses of inducer
$I_1= 0.0, 3.9\times10^{-4}, 1.6\times10^{-3}, 6.3\times10^{-3}, 2.5\times 10^{-2}, 
0.1, 0.4, 1.6, 6.4, 12.8$ mM, with two different reporters ({\it gfp}-30 and 
{\it gfp}-34). See Tables~\ref{tab:gfp30-data} and~\ref{tab:gfp34-data} 
Instead of assuming equilibrium and using only the data for $t > 300$ min as 
done previously \citep{Wang2011}, we apply Algorithm~\ref{alg:method} to all the data 
with the full dynamical equation~(\ref{eq:GFP}) to estimate 
\mbox{$\btheta = [\alpha, k_1, n_1, K_1, d]$}. In this 
case, we used initial priors $U(0,1)$ for $\alpha$ and $n_1$; and $U(0, 20)$ for $k_1$, 
$K_1$ and $d$. The other parameters were set to:
 $J=1000$, $B=50$, $p_m=0.95$, and $Tol=10^{-5}$.

Our algorithm converged after five iterations to the parameter values in 
Table~\ref{tab:GFP-data}. The parameter estimates provide good fits to both 
the time courses (Fig.~\ref{fig:GFP-panel}B) and to the dose response data 
(Fig.~\ref{fig:GFP-panel}A). The values of $K_1^\ddagger$ and $n_1^\ddagger$ 
obtained here are similar those obtained in \cite{Wang2010} by using only 
stationary data. 
This is reassuring  since these parameters are related to the dose threshold to 
half maximal response and  to the steepness of the sigmoidal response, both static properties. 
On the other hand,  the values of $\alpha$ and the ratio $k_1/d$ differ to some 
extent due to the (imperfect) assumption in \cite{Wang2010} that steady state had 
been reached at $t=300$ min. As Fig.~\ref{fig:GFP-panel}B shows, $G$ is not 
at steady state then. Hence the parameter values obtained with our method should 
give a more faithful representation of the true dynamical response of the switch.

\begin{table}[!tpb]
  \begin{center}
   {\begin{tabular}{c|cc|cc} \hline
   & \multicolumn{2}{c|}{\cite{Wang2010}} & 
       \multicolumn{2}{c}{Algorithm~\ref{alg:method}} \\ \hline
       Parameter & {\it gfp}-30 & {\it gfp}-34 & {\it gfp}-30 & {\it gfp}-34 \\ \hline
        $\alpha^\ddagger$ & $0.0012\pm 0.027$ & $1.4720\times 10^{-9}$ & 0.0043 & 0.0024 \vspace{1mm} \\
        $k_1^\ddagger$    &  N/A & N/A & 76.1354 & 63.6650 \vspace{1mm}\\
        $n_1^\ddagger$    &  $1.3700\pm 0.270$ & $1.3690\pm 0.021$ & 1.4832 & 1.3879 \vspace{1mm}\\
        $K_1^\ddagger$    &  $0.2280\pm 0.039$ & $0.2590\pm 0.021$ & 0.2467 & 0.2641\vspace{1mm} \\
        $d^\ddagger$      & N/A & N/A & 0.0069 & 0.0052 \\
        $k_1^\ddagger/d^\ddagger$  & $9456\pm 487$ & $7648\pm152$ & 10983.34 & 12163.04 \\
       \hline
   \end{tabular}}
   \caption{{\small Parameter values obtained from {\it gfp}-30 and {\it gfp}-34 data. 
  In \cite{Wang2010}, only the steady state solution was used. Hence only the ratio of $k_1$ and $d$ can be estimated. }}
   \label{tab:GFP-data}
  \end{center}
\end{table}

\section{Discussion}
\label{sec:Discussion}

In this work, we have presented an optimisation algorithm that brings together 
ingredients from Evolutionary Algorithms, local optimisation and Sequential Monte 
Carlo. The method is particularly useful for determining parameters of 
ordinary differential equation models from data. Our approach can also be used in 
other contexts where an optimisation problem has to be solved on complex landscapes, 
or when the objective function cannot be written explicitly. 
The algorithm proceeds by generating a population of solutions through Monte Carlo 
sampling from a prior distribution and refining those solutions through a 
combination of local optimisation and culling. A new prior is then created as a 
mixture of a historical prior (which records the broadest possible range of 
solutions found) and the distribution of the optimised population. This iterative 
process combines a strong concentration of the Monte Carlo sampling through local
optimisation with the possibility that solutions can be found outside of 
the initial prior.

We have illustrated the application of the algorithm to ODE models of 
biological interest and have found it to perform efficiently. The algorithm also works well 
when applied to larger problems with tens of parameters in a signal transduction model 
(paper in preparation).  The efficiency of the algorithm hinges on selecting 
appropriate running parameters. For instance, the number of samples from the prior $J$
should be large enough to allow for significant sampling of the parameter space 
while small enough to limit the computational cost. We have found that simulating 
$J=350-500$ points in models of up to 10 parameters and keeping the best 15\% 
of the local minima leads to termination within fewer than 20 iterations.  
In our implementation, the Nelder-Mead minimisation is capped at 300 evaluations. 
These guidelines would result in  ~150,000 evaluations of the objective 
function per iteration. Therefore our method can become computationally costly 
if the objective function is expensive to evaluate, 
e.g. in stiff models that are difficult to solve numerically.
In essence, our algorithm proposes a trade-off: fewer but more 
costly iterations. It is important to remark that,  as with any other 
optimisation heuristic for non convex problems, there are no strict guarantees of 
convergence to the global minimum. Therefore, it is always advisable to run the 
method with different starting points and different settings to check for 
consistency of the solutions obtained.

The generation of iterative samples of the parameters draws inspiration from Monte 
Carlo methods \citep{Sisson2007, Toni2009, Toni2009a} but without pursuing the strict
guarantees that the nested structure of the distributions in ABC-SMC provides.
Our evolutionary approach adopts a highly focused Monte Carlo sampling
driven by a sharp local search with culling. Hence our iterative procedure 
generates samples that only reflect properties of the set of local minima 
(up to numerical cutoffs) without any focus on the global convergence of 
the distributions.
As noted by \cite{Toni2009a}, the distributions of the parameters 
(both their sequence and the final distributions) give information
about the sensitivity of the parameters: parameters with 
narrow support will be more sensitive than those with wider support.
Future developments of the method will focus on establishing a suitable theoretical 
framework that facilitates its use in model selection. 
Other work will consider the possibility of incorporating a stochastic 
ranking strategy in the selection of solutions, similar to that 
present in the SRES algorithm \citep{Runarsson2000}, in order to solve more general
optimisation problems with complex feasible regions.


\section*{Acknowledgements}
The authors would like to thank C.~Barnes, T.~Ellis, E.~Gardu\~no, H.~Harrington, 
M.~Owen, M.~Stumpf, and S.~Yaliraki for their comments and suggestions.
\paragraph{Funding:} MBD is supported by a BBSRC-Microsoft 
Research Dorothy Hodgkin Postgraduate Award. This work was partly supported by the 
US Office of Naval Research, and by BBSRC through LoLa grant BB/G020434/1 (MB), and EPSRC 
through grant EP/I017267/1 under the 
\textit{Mathematics underpinning the Digital Economy} program (MB).

\appendix

\begin{small}

\section{BPM model data}

\begin{table}[htp] 
  \begin{center}
   {\center \begin{tabularx}{86mm}{|X|X|} \hline
      t & R \\ \hline
      0 & 0 \\
      20 & 43.5373 \\
      40 & 13.3667 \\
      60 & 140.8903 \\
      80 & 29.2816 \\
      100 & 108.1722 \\
      120 & 19.0093 \\
      140 & 75.0065 \\
      160 & 14.4018 \\
      180 & 50.4473 \\
      200 & 217.1082 \\
      \hline
  \end{tabularx}}
   \caption{BPM data.}
   \label{tab:BPM-data}
\end{center}
\end{table}

Table~\ref{tab:BPM-data} shows data obtained from a simulation of the BPM model from 
equations~(6) using parameters $\alpha = 240$ and $\beta = 0.15$, initial conditions 
$R(0)=0$, $E(0)=0$, $P(0)=0$, and adding random noise sampled from a $N(0, 15^2)$
distribution. Only the data for variable $R$ was obtained.

\section{SIR model data}

\begin{table}[htp]
\begin{center}
     {\begin{tabularx}{86mm}{|X|XXX|} \hline
         t & S & I & R \\ \hline
         0.6 & 0.12 & 13.17 & 9.42 \\
         1.0 & 0.12 & 7.17 & 11.19 \\
         2.0 & 0.10 & 2.36 & 10.04 \\
         3.0 & 0.38 & 0.92 & 6.87 \\
         4.0 & 1.00 & 0.62 & 4.45 \\
         5.0 & 1.20 & 0.17 & 3.01 \\
         6.0 & 1.46 & 0.28 & 1.76 \\
         7.0 & 1.38 & 0.10 & 1.29 \\
         8.0 & 1.57 & 0.03 & 0.82 \\
         9.0 & 1.46 & 0.29 & 0.52 \\
         10.0 & 1.25 & 0.10 & 0.23 \\
         11.0 & 1.56 & 0.22 & 0.20 \\
         \hline
     \end{tabularx}}
     \caption{SIR data.}
     \label{tab:SIR-data}
\end{center}
\end{table}

Table~\ref{tab:SIR-data} shows data for the SIR model generated from equations~(8)
using initial conditions $S(0)=20$, $I(0)=10$, and $R(0)=0$ with added random noise 
sampled from a $N(0,0.2^2)$ distribution as appears in Ref.~\cite{Toni2009a}.

\section{Genetic switch data}

Tables~\ref{tab:gfp30-data} and~\ref{tab:gfp34-data} show the fluorescent response
of IPTG-induced genetic switches described in Ref.~\cite{Wang2010} 
and~\cite{Wang2011}.

      \begin{table}[ht]
      \begin{center} 
{\small \begin{tabular}{|c|c|c|c|c|c|c|c|c|c|c|c|} \hline
     t  & 0mM & 0.0004mM & 0.0016mM & 0.0063mM & 0.025mM & 0.1mM & 0.4mM 
    & 1.6mM & 6.4mM & 12.8mM \\ \hline
          0 & 0 & 0 & 0 & 0 & 0 & 0 & 0 & 0 & 0 & 0 \\
   140 & 88.6 & 177.8 & 174.4 & 197.8 & 210.4 & 1043.6 & 3945.8 & 5971 & 6643.8 & 6521.8 \\
   160 & 120.2 & 156.4 & 160.6 & 165.6 & 209.8 & 1300.8 & 4695.2 & 6768.4 & 7361.8 & 7513.8 \\
   180 & 66.6 & 96.4 & 94.6 & 126.4 & 171.6 & 1438.4 & 5238.8 & 7465.2 & 7801 & 8002.4 \\
   200 & 42.8 & 72.2 & 76.2 & 88 & 134.2 & 1578 & 5658 & 7914 & 8458 & 8542.8 \\
   220 & 37 & 64.8 & 61.2 & 55 & 135.8 & 1667 & 5799.6 & 8380.2 & 8976 & 8914.8 \\
   240 & 39.6 & 56.6 & 60.4 & 65.8 & 142.8 & 1758.6 & 6108.6 & 8601.4 & 9172.6 & 8957 \\
   260 & 36.2 & 47.6 & 62 & 69.8 & 143.6 & 1859.8 & 6104 & 9041.8 & 9528.6 & 9252.8 \\
   280 & 50.8 & 55.6 & 58.2 & 74.2 & 170.6 & 1968.2 & 6554.4 & 9071.6 & 9449 & 9018.4 \\
   300 & 39.6 & 51 & 40.8 & 60.2 & 197.8 & 2143.4 & 6452.2 & 8396.2 & 9269.2 & 9261.2 \\
   320 & 50.4 & 62.8 & 65.6 & 82 & 273.6 & 2317.8 & 6880.8 & 8941.2 & 9887.6 & 9982.8 \\
   340 & 53.8 & 71.4 & 71 & 88.6 & 296 & 2512.8 & 7052.2 & 8972.8 & 9694.6 & 10108 \\
   360 & 45.6 & 66 & 61.6 & 69.2 & 340.8 & 2639.2 & 7047.8 & 9103.6 & 9911 & 10018.4 \\
   \hline
\end{tabular}}\\
{\small \begin{tabular}{|c|c|c|c|c|c|c|c|c|c|c|c|} \hline
     t  & 0mM & 0.0004mM & 0.0016mM & 0.0063mM & 0.025mM & 0.1mM & 0.4mM 
    & 1.6mM & 6.4mM & 12.8mM \\ \hline
      0 & 0 & 0 & 0 & 0 & 0 & 0 & 0 & 0 & 0 & 0 \\
   140 & 215 & 163.4 & 124.8 & 134 & 119 & 230.4 & 721.2 & 1001.8 & 1095.8 & 701 \\
   160 & 141.6 & 116.6 & 95.4 & 86 & 40 & 320.6 & 937 & 1112.2 & 1054 & 903.2 \\
   180 & 131.6 & 112.2 & 117.6 & 84 & 81 & 252.2 & 825.2 & 727.4 & 1026.8 & 679.2 \\
   200 & 69.8 & 42.4 & 37.8 & 39 & 44.2 & 225.2 & 688.4 & 829.8 & 761.6 & 584.6 \\
   220 & 55 & 58.4 & 59 & 60.6 & 50.4 & 169.2 & 645.8 & 713.6 & 739.6 & 454 \\
   240 & 38.8 & 48 & 30.8 & 43.4 & 42.2 & 148.8 & 366 & 418.6 & 453.8 & 668.2 \\
   260 & 42.2 & 44 & 48.6 & 41 & 53.8 & 152.8 & 496.4 & 638.4 & 547.8 & 626.2 \\
   280 & 55.2 & 54.4 & 51.8 & 53.6 & 76 & 257.2 & 498.2 & 722.2 & 889.8 & 606.2 \\
   300 & 50.4 & 57.4 & 62 & 67.8 & 95 & 339.8 & 447.4 & 835.6 & 693.2 & 602.6 \\
   320 & 52.6 & 69.6 & 78.4 & 81.2 & 146.8 & 385.8 & 540.4 & 776.4 & 1084.2 & 580 \\
   340 & 57 & 60.6 & 73.8 & 65.6 & 144.6 & 401.2 & 466.4 & 396.6 & 560.4 & 702 \\
   360 & 61.6 & 73.2 & 77.2 & 68.6 & 151 & 400 & 374.8 & 251 & 742 & 436.2 \\

   \hline
\end{tabular}}
\caption{{\it gfp}30 fluorescence measurements (top) and standard deviations (bottom).}
\label{tab:gfp30-data}
\end{center}
\end{table}

      \begin{table}[ht]
      \begin{center} 
{\small \begin{tabular}{|c|c|c|c|c|c|c|c|c|c|c|c|} \hline
     t  & 0mM & 0.0004mM & 0.0016mM & 0.0063mM & 0.025mM & 0.1mM & 0.4mM 
    & 1.6mM & 6.4mM & 12.8mM \\ \hline
     0 & 0 & 0 & 0 & 0 & 0 & 0 & 0 & 0 & 0 & 0 \\
   140 & 149.1 & 199.7 & 107.4 & 124.6 & 242.4 & 801.9 & 2682.7 & 4292.3 & 4633.3 & 4923.8 \\
   160 & 96 & 212.2 & 121.6 & 78.4 & 199.3 & 945 & 3192.9 & 4893.7 & 5243.3 & 5572.6 \\
   180 & 64.3 & 178.7 & 73.7 & 40.4 & 158.7 & 1083.8 & 3598.4 & 5362.7 & 5762.6 & 6139.4 \\
   200 & 32.2 & 92.5 & 43.2 & 43.7 & 135.1 & 1190.5 & 3961.4 & 5901.6 & 6282.9 & 6499.9 \\
   220 & 56.4 & 86.5 & 51.5 & 43.5 & 142.8 & 1320.4 & 4274.4 & 6218.6 & 6589.5 & 6866.5 \\
   240 & 42.4 & 54.6 & 16.5 & 23.9 & 116.3 & 1330.6 & 4424.9 & 6247.9 & 6514.3 & 6815.1 \\
   260 & 31 & 49.9 & 11.3 & 13.4 & 100.4 & 1422.8 & 4583.5 & 6531 & 6917.5 & 7177.6 \\
   280 & 34.7 & 55.5 & 13 & 16.4 & 107.1 & 1535.8 & 4680.4 & 6609.6 & 7247.2 & 7290.1 \\
   300 & 33.2 & 46.1 & 21.7 & 22.1 & 129.7 & 1675.5 & 4958.5 & 6949.3 & 7620.3 & 7631.3 \\
   320 & 29.5 & 39 & 8.7 & 22.5 & 154 & 1824.5 & 5122.3 & 7053.4 & 7642.7 & 7645.1 \\
   340 & 31.2 & 43.2 & 19.1 & 27.1 & 172.2 & 1836.2 & 5282.7 & 7156.9 & 7661.2 & 7889.3 \\
   360 & 28 & 40 & 10.9 & 28.9 & 202.4 & 1979.3 & 5456.4 & 7245.6 & 7899.1 & 7910.6 \\
   \hline
\end{tabular}}\\
{\small \begin{tabular}{|c|c|c|c|c|c|c|c|c|c|c|c|} \hline
     t  & 0mM & 0.0004mM & 0.0016mM & 0.0063mM & 0.025mM & 0.1mM & 0.4mM 
    & 1.6mM & 6.4mM & 12.8mM \\ \hline
     0 & 0 & 0 & 0 & 0 & 0 & 0 & 0 & 0 & 0 & 0 \\
   140 & 89.4 & 85.8 & 209.2 & 120.8 & 77.8 & 175.8 & 383.6 & 295.4 & 332.6 & 382 \\
   160 & 59.4 & 23 & 166.4 & 111.6 & 40.6 & 188.8 & 572.2 & 391.6 & 430.6 & 326.2 \\
   180 & 31.6 & 38.6 & 135.4 & 51.2 & 24.8 & 210.6 & 597 & 370.8 & 467.6 & 363.8 \\
   200 & 45.2 & 60.4 & 83.2 & 65.2 & 42 & 166 & 573.2 & 273.6 & 341.6 & 337 \\
   220 & 14 & 27.4 & 90.2 & 51.2 & 25 & 90 & 513.8 & 249.6 & 234 & 145.2 \\
   240 & 25.2 & 32.2 & 53.8 & 30.6 & 16.2 & 70 & 475.2 & 187.6 & 464.8 & 168 \\
   260 & 14.8 & 17.2 & 47.4 & 23.8 & 14.2 & 68.8 & 511.8 & 256 & 300.6 & 214 \\
   280 & 20 & 15.4 & 46.6 & 16.6 & 15.8 & 70.6 & 395.8 & 237.6 & 313.6 & 454.6 \\
   300 & 17.8 & 17.8 & 37.8 & 29.8 & 29.2 & 178.2 & 486.6 & 383.8 & 416.2 & 377.2 \\
   320 & 21 & 21.2 & 43 & 26.4 & 46.8 & 216.2 & 519.6 & 507.4 & 674.8 & 227 \\
   340 & 26 & 22.2 & 36.8 & 25.4 & 46.6 & 340.8 & 495.6 & 655.6 & 594.2 & 299.4 \\
   360 & 15.2 & 13 & 38.4 & 8.6 & 50 & 350.4 & 604.8 & 434.2 & 853.8 & 387.8 \\
   \hline
\end{tabular}}
\caption{{\it gfp}34 fluorescence measurements (top) and standard deviations (bottom).}
\label{tab:gfp34-data}
\end{center}
\end{table}

\end{small}


\newpage

\begin{thebibliography}{}

\bibitem[Alon(2007)Alon]{Alon2007}
Alon, U. (2007).
\newblock {\em {An introduction to systems biology: design principles of
  biological circuits}\/}.
\newblock Chapman and Hall/CRC mathematical \& computational biology series.
  Chapman \& Hall/CRC.

\bibitem[Anderson and May(1992)Anderson and May]{Anderson1992}
Anderson, R.~M. and May, R.~M. (1992).
\newblock {\em {Infectious Diseases of Humans Dynamics and Control}\/}.
\newblock Oxford University Press.

\bibitem[Bliss {\em et~al.}(1982)Bliss, Painter, and Marr]{Bliss1982}
Bliss, R.~D., Painter, P.~R., and Marr, A.~G. (1982).
\newblock {Role of feedback inhibition in stabilizing the classical operon}.
\newblock {\em Journal of Theoretical Biology\/}, {\bf 97}(2), 177 -- 193.

\bibitem[Brewer {\em et~al.}(2008)Brewer, Barenco, Callard, Hubank, and
  Stark]{Brewer2008}
Brewer, D., Barenco, M., Callard, R., Hubank, M., and Stark, J. (2008).
\newblock {Fitting ordinary differential equations to short time course data}.
\newblock {\em Philosophical Transactions of the Royal Society A: Mathematical,
  Physical and Engineering Sciences\/}, {\bf 366}(1865), 519--544.

\bibitem[Brown and Sethna(2003)Brown and Sethna]{Brown2003}
Brown, K.~S. and Sethna, J.~P. (2003).
\newblock {Statistical mechanical approaches to models with many poorly known
  parameters}.
\newblock {\em Physical Review E\/}, {\bf 68}(2), 021904.

\bibitem[Chen {\em et~al.}(2010)Chen, Niepel, and Sorger]{Chen2010}
Chen, W.~W., Niepel, M., and Sorger, P.~K. (2010).
\newblock {Classic and contemporary approaches to modeling biochemical
  reactions.}
\newblock {\em Genes Dev\/}, {\bf 24}(17), 1861--1875.

\bibitem[Edelstein-Keshet(1988)Edelstein-Keshet]{Edelstein-Keshet1988}
Edelstein-Keshet, L. (1988).
\newblock {\em {Mathematical models in biology}\/}.
\newblock Classics in applied mathematics. SIAM.

\bibitem[Gershenfeld(1999)Gershenfeld]{Gershenfeld1999}
Gershenfeld, N. (1999).
\newblock {\em {The nature of mathematical modeling}\/}.
\newblock Cambridge University Press.

\bibitem[Gutenkunst {\em et~al.}(2007)Gutenkunst, Waterfall, Casey, Brown,
  Myers, and Sethna]{Gutenkunst2007}
Gutenkunst, R.~N., Waterfall, J.~J., Casey, F.~P., Brown, K.~S., Myers, C.~R.,
  and Sethna, J.~P. (2007).
\newblock {Universally Sloppy Parameter Sensitivities in Systems Biology
  Models}.
\newblock {\em PLoS Comput Biol\/}, {\bf 3}(10), e189.

\bibitem[Kirkpatrick {\em et~al.}(1983)Kirkpatrick, Gelatt, and
  Vecchi]{Kirkpatrick1983}
Kirkpatrick, S., Gelatt, C.~D., and Vecchi, M.~P. (1983).
\newblock {Optimization by simulated annealing.}
\newblock {\em Science\/}, {\bf 220}(4598), 671--680.

\bibitem[Lawson and Hanson(1995)Lawson and Hanson]{Lawson1995}
Lawson, C. and Hanson, R. (1995).
\newblock {\em {Solving least squares problems}\/}.
\newblock Classics in applied mathematics. SIAM.

\bibitem[Mitchell(1997)Mitchell]{Mitchell1997}
Mitchell, T.~M. (1997).
\newblock {\em {Machine Learning}\/}.
\newblock McGraw-Hill, New York.

\bibitem[Moles {\em et~al.}(2003)Moles, Mendes, and Banga]{Moles2003}
Moles, C.~G., Mendes, P., and Banga, J.~R. (2003).
\newblock {Parameter Estimation in Biochemical Pathways: A Comparison of Global
  Optimization Methods}.
\newblock {\em Genome Research\/}, {\bf 13}(11), 2467--2474.

\bibitem[Nelder and Mead(1965)Nelder and Mead]{Nelder1965}
Nelder, J.~A. and Mead, R. (1965).
\newblock {A Simplex Method for Function Minimization}.
\newblock {\em The Computer Journal\/}, {\bf 7}(4), 308--313.

\bibitem[Papachristodoulou and Recht(2007)Papachristodoulou and
  Recht]{Papachristodoulou2007}
Papachristodoulou, A. and Recht, B. (2007).
\newblock {Determining Interconnections in Chemical Reaction Networks}.
\newblock In {\em American Control Conference, 2007.}, pages 4872--4877.

\bibitem[Powell(1998)Powell]{Powell1997}
Powell, M. J.~D. (1998).
\newblock {Direct Search Algorithms for Optimization Calculations}.
\newblock {\em Acta Numerica\/}, {\bf 7}(-1), 287--336.

\bibitem[Runarsson and Yao(2000)Runarsson and Yao]{Runarsson2000}
Runarsson, T. and Yao, X. (2000).
\newblock {Stochastic ranking for constrained evolutionary optimization}.
\newblock {\em IEEE Transactions on Evolutionary Computation\/}, {\bf 4}(3),
  284 --294.

\bibitem[Schwefel(1995)Schwefel]{Schwefel1995}
Schwefel, H. (1995).
\newblock {\em {Evolution and optimum seeking}\/}.
\newblock Sixth-generation computer technology series. Wiley.

\bibitem[Sisson {\em et~al.}(2007)Sisson, Fan, and Tanaka]{Sisson2007}
Sisson, S.~A., Fan, Y., and Tanaka, M.~M. (2007).
\newblock {Sequential Monte Carlo without likelihoods.}
\newblock {\em Proc Natl Acad Sci USA\/}, {\bf 104}(6), 1760--1765.

\bibitem[Sontag(2002)Sontag]{Sontag2002}
Sontag, E. (2002).
\newblock {For Differential Equations with r Parameters, 2r+1 Experiments Are
  Enough for Identification}.
\newblock {\em Journal of Nonlinear Science\/}, {\bf 12}, 553--583.

\bibitem[Strogatz(1994)Strogatz]{Strogatz1994}
Strogatz, S.~H. (1994).
\newblock {\em {Nonlinear Dynamics And Chaos. With Applications to Physics,
  Biology, Chemistry, and Engineering}\/}.
\newblock Studies in nonlinearity. Perseus Books Group.

\bibitem[Szallasi {\em et~al.}(2006)Szallasi, Stelling, and
  Periwal]{Szallasi2006}
Szallasi, Z., Stelling, J., and Periwal, V. (2006).
\newblock {\em {System modeling in cell biology: from concepts to nuts and
  bolts}\/}.
\newblock Bradford Book. MIT Press.

\bibitem[Toni and Stumpf(2009)Toni and Stumpf]{Toni2009}
Toni, T. and Stumpf, M. P.~H. (2009).
\newblock {Simulation-based model selection for dynamical systems in systems
  and population biology}.
\newblock {\em Bioinformatics\/}.

\bibitem[Toni {\em et~al.}(2009)Toni, Welch, Strelkowa, Ipsen, and
  Stumpf]{Toni2009a}
Toni, T., Welch, D., Strelkowa, N., Ipsen, A., and Stumpf, M.~P. (2009).
\newblock {Approximate Bayesian computation scheme for parameter inference and
  model selection in dynamical systems}.
\newblock {\em Journal of The Royal Society Interface\/}, {\bf 6}(31),
  187--202.

\bibitem[Wang(2010)Wang]{Wang2010}
Wang, B. (2010).
\newblock {\em {Design and Functional Assembly of Synthetic Biological Parts
  and Devices}\/}.
\newblock Ph.D. thesis, Imperial College London.

\bibitem[Wang {\em et~al.}(2011)Wang, Kitney, Joly, and Buck]{Wang2011}
Wang, B., Kitney, R.~I., Joly, N., and Buck, M. (2011).
\newblock {Engineering modular and orthogonal genetic logic gates for robust
  digital-like synthetic biology.}
\newblock {\em Nat Commun\/}, {\bf 2}, 508.

\bibitem[Yates {\em et~al.}(2001)Yates, Chan, Callard, George, and
  Stark]{Yates2001}
Yates, A., Chan, C. C.~W., Callard, R.~E., George, A. J.~T., and Stark, J.
  (2001).
\newblock {An approach to modelling in immunology}.
\newblock {\em Brief Bioinform\/}, {\bf 2}(3), 245--257.

\bibitem[Zi and Klipp(2006)Zi and Klipp]{Zi2006}
Zi, Z. and Klipp, E. (2006).
\newblock {SBML-PET: a Systems Biology Markup Language-based parameter
  estimation tool}.
\newblock {\em Bioinformatics\/}, {\bf 22}(21), 2704--2705.

\end{thebibliography}
\bibliographystyle{natbib}

\end{document}